\documentclass[fleqn,usenatbib]{mnras}

\usepackage{newtxtext,newtxmath}
\usepackage[T1]{fontenc}
\usepackage{ae,aecompl}

\usepackage{graphicx}	% Including figure files
\usepackage{amsmath}	% Advanced maths commands
\usepackage{amssymb}	% Extra maths symbols
\usepackage{xspace}        %  For problem with missing space in using newcommand

\newcommand{\target}{IRAS F11119$+$3257}

\title[Jet in \target{}]{A two-sided but significantly beamed jet in the supercritical accretion quasar \target{}}

\author[J. Yang et al.]
{Jun Yang,$^{1}$\thanks{E-mail: jun.yang@chalmers.se}
Zsolt Paragi,$^{2}$
Tao An,$^{3}$
Willem A. Baan,$^{4, 5}$
Prashanth Mohan$^{3}$
\and and Xiang Liu$^{5}$
\\
\\
$^{1}$Department of Space, Earth and Environment, Chalmers University of Technology, Onsala Space Observatory, SE-439 92 Onsala, Sweden \\
$^{2}$Joint Institute for VLBI ERIC (JIVE), Postbus 2, NL-7990 AA Dwingeloo, the Netherlands \\
$^{3}$Shanghai Astronomical Observatory, Key Laboratory of Radio Astronomy, Chinese Academy of Sciences, 200030 Shanghai, China \\
$^{4}$Netherlands Institute for Radio Astronomy ASTRON, NL-7991 PD Dwingeloo, the Netherlands \\
$^{5}$Xinjiang Astronomical Observatory, Key Laboratory of Radio Astronomy, Chinese Academy of Sciences, 150 Science 1-Street, 830011 Urumqi, China \\
}
\date{Accepted 2020 XXX. Received 2020 YYY; in original form 2020 ZZZ}

\pubyear{2020}

\begin{document}
\label{firstpage}
\pagerange{\pageref{firstpage}--\pageref{lastpage}}
\maketitle

% <=250 words
\begin{abstract}
Highly accreting quasars are quite luminous in the X-ray and optical regimes. While, they tend to become radio quiet and have optically thin radio spectra. Among the known quasars, \target{} is a supercritical accretion source because it has a bolometric luminosity above the Eddington limit and extremely powerful X-ray outflows. To probe its radio structure, we investigated its radio spectrum between 0.15 and 96.15 GHz and performed very-long-baseline interferometric (VLBI) observations with the European VLBI Network (EVN) at 1.66 and 4.93 GHz. The deep EVN image at 1.66 GHz shows a two-sided jet with a projected separation about two hundred parsec and a very high flux density ratio of about 290. Together with the best-fit value of the integrated spectral index of $-$1.31$\pm$0.02 in the optically thin part, we infer that the approaching jet has an intrinsic speed at least 0.57 times of the light speed. This is a new record among the known all kinds of super-Eddington accreting sources and unlikely accelerated by the radiation pressure. We propose a scenario in which \target{} is an unusual compact symmetric object with a small jet viewing angle and a radio spectrum peaking at 0.53$\pm$0.06 GHz mainly due to the synchrotron self-absorption. 

\end{abstract}


%%%%%%%%%%%%%%%%%%%%%%%%%%%%%%%%%%%%%%%%%%%%%%%%%%

%%%%%%%%%%%%%%%%% BODY OF PAPER %%%%%%%%%%%%%%%%%%

\section{Introduction}
\label{sec1}
Nearby active galactic nuclei (AGN) hosting highly accreting (accretion rate comparable to or exceeding the Eddington rate) supermassive black holes (SMBHs) are useful targets for monitoring observations in several contexts. Some of these include studies of powerful X-ray absorption outflows and their impact on the host galaxy evolution \citep[e.g.][]{Kormendy2013, Nardini2015, Tombesi2015, Tombesi2016}, and in exploring radio jet and outflow activity in an extreme accretion regime \citep[e.g.][]{Giroletti2017, Yang2018}. These highly-accreting AGN are typically radio-quiet \citep[e.g.][]{Greene2006, Panessa2007, Sikora2007} with a steep radio spectrum \citep{Laor2019} when their accretion rates approach or exceed the Eddington limit. The existence of young, possibly episodic jets moving at mildly relativistic jet speeds in these AGN \citep[e.g.][]{Panessa2019} thus remain debatable. Further, their high redshift counterparts are probes of rapid SMBH growth \citep[e.g.][]{Volonteri2015}, feedback interaction with the host galaxy through jet and radiative processes \citep[e.g.][]{Pacucci2015}, and are used as standard candles for determining cosmological luminosity distance \citep[e.g.][]{Wang2013, Marziani2014}.

The quasar \target{} (J1114$+$3241, B1111$+$329) at a redshift $z = 0.189$ is a type 1 ultra-luminous infra-red galaxy (ULIRG) and hosts strong molecular outflows, with the emission dominated by the AGN component \citep{Veilleux2013, Veilleux2017}. Based on a SMBH mass of $M_{\rm bh} \approx 2 \times 10^7$ M$_{\sun}$ calibrated for a sample of similar ULIRG sources \citep{Kawakatu2007}, the bolometric luminosity $L_{\rm b} = 5 L_{\rm Edd}$, where $L_{\rm Edd}$ is the Eddington luminosity \citep{Tombesi2015}. Using a correlation relation between infra-red and radio luminosities for starburst galaxies, it is found that for \target{} the AGN component far exceeds the starburst contribution \citep{Komossa2006}. In addition to the molecular outflows, it also hosts wide-aperture energetic radiation driven X-ray emitting winds, suggesting a likely energy conserving quasar-mode feedback \citep{Tombesi2015, Tombesi2017}. 

The quasar \target{} has a relatively bright radio counterpart. The early survey of the Bologna Northern Cross Radio Telescope (BNCRT) at 408~MHz \citep{Colla1970} found a half-Jy radio counterpart. Later, it was observed by more radio telescopes at multiple frequencies. The radio flux densities reported in literature over the past 30 years are listed in Table~\ref{tab1}. All these existing observations indicate a compact emission structure \citep[e.g.][]{Berton2018, Veilleux2017}. Although it has an increasingly steep radio spectrum at frequencies $\ga$1~GHz, it may not be classified as radio-quiet owing to a high radio power ($\approx 10^{25}$~W~Hz$^{-1}$) reminiscent of a radio-loud quasar \citep{Komossa2006}. 

The radio jet, radiation pressure from the disk emission and disk winds can contribute to driving the powerful X-ray outflows in this source \citep[e.g.][]{Tombesi2016}. To probe this and to possibly resolve the compact but steep spectrum radio source, we conducted high resolution VLBI observations of the pc-scale region. This paper is compiled as follows. In Section~\ref{sec2}, we describe the radio observations of \target{} and the data reduction. In Section~\ref{sec3}, we report the broad-band radio spectrum and the radio morphology. In Section~\ref{sec4}, we discuss the radio core, constraints on the jet parameters and the relation to young radio sources. In Section~\ref{sec5}, we present the main conclusions from our study. Throughout the paper, a standard $\Lambda$CDM cosmological model with H$_\mathrm{0}$~=~71~km~s$^{-1}$~Mpc$^{-1}$, $\Omega_\mathrm{m}$~=~0.27, $\Omega_{\Lambda}$~=~0.73 is adopted; the VLBI images then have a scale of 3.9~pc mas$^{-1}$.

\begin{table}
\caption{List of the radio flux densities of \target{}.  The used radio facilities are the Giant Metrewave Radio Telescope (GMRT), the Westerbork Synthesis Radio Telescope (WSRT),  the Robert C. Byrd Green Bank Telescope (GBT), the Karl G. Jansky Very Large Array (VLA), and the Atacama Large Millimeter Array (ALMA). }
\label{tab1}
\centering
\begin{tabular}{cccc}
\hline\hline
 Freq.        & Flux              &  Array    &   Reference          \\
 (GHz)       & (mJy)            &                     &                             \\
\hline
0.15          & 154$\pm$16 & GMRT & \citet{Intema2017} \\  
0.33          & 241$\pm$24 & WSRT  & \citet{Rengelink1997} \\   
1.40          &105$\pm$5    & VLA      & \citet{Becker1995} \\
1.40          & 110$\pm$3    & VLA      & \citet{Condon1998} \\
1.43          & 106$\pm$5    & VLA      & This paper            \\
4.85          & 23$\pm$4      & GBT     & \citet{Gregory1996} \\
5.23          & 21.0$\pm$1.1 & VLA    & \citet{Berton2018}  \\
8.44          & 11$\pm$0.6   &  VLA     & This paper             \\
14.94        & 4.6$\pm$0.3  & VLA     & This paper              \\
96.15        & 0.5$\pm$0.05  & ALMA  &  \citet{Veilleux2017} \\

\hline
\end{tabular}
\end{table}

\section{Observations and data reduction}
\label{sec2}
\subsection{The EVN experiments at 1.66 and 4.93 GHz}
\label{sec2-1}
We observed \target{} with the EVN at 1.66 and 4.93~GHz in 2016. The experiment setups are summarised in Table~\ref{tab2}. The participating telescopes were Robledo (Ro), Sardinia (Sr), Hartesbeesthoek (Hh), Zelenchukskaya (Zc), Sveltoe (Sv), Torun (Tr), Urumqi (Ur), Tianma (T6), Onsala (O8), Medicina (Mc), Effelsberg (Ef), Westerbork (Wb, single dish), Jodrell Bank Lovell (Jb1) and Mk2 (Jb2). The correlation was done by the EVN software correlator \citep[SFXC,][]{Keimpema2015} at JIVE (Joint Institute for VLBI, ERIC) using the typical correlation parameters. 

The short EVN observations at 4.93 GHz were performed in the e-VLBI mode on 2016 November 16. The data rate was reduced to 1024 Mbps for T6 due to the network limitation, and Tr and Jb2 due to the VLBI backends. To calibrate the data and measure a precise position for our faint target \target{}, a bright compact source J1111$+$3252 \citep{Helmboldt2007}, about 41 arcmin apart from \target{}, was also observed periodically. The phase-referencing calibrator position is RA~$=11^{\rm h}11^{\rm m}31\fs77219$, Dec.~$=32\degr52\arcmin55\farcs7847$ (J2000, $\sigma_{\rm ra}=\sigma_{\rm dec}=0.16$ mas) in the source catalogue\footnote{\url{http://astrogeo.org/vlbi/solutions/rfc_2019d/}} 
provided by L. Petrov from the Goddard Space Flight Centre VLBI group. The calibrator position has an offset of 1.4 mas with respect to the sub-mas-precision optical position in the second data release \citep[DR2,][]{Brown2018} of the \textit{Gaia} mission \citep{Prusti2016}. The nodding observations used a cycle period of about six minutes (1.5~min for J1111$+$3252, 4~min for \target{}). All the telescopes had an elevation of $\geq$18~deg during the observations. 

The full EVN observations at 1.66 GHz were carried out in the disk-recording mode on 2016 March 8. There were thirteen stations participating the four-hour observations. Because the target source is relatively bright at 1.66 GHz, the phase-referencing calibrator was not observed. This allowed us to reach an on-target time of about 200 min.

The visibility data were calibrated using the National Radio Astronomy Observatory (NRAO) software package Astronomical Image Processing System \citep[\textsc{aips}, ][]{Greisen2003}. As the digital filters provided by the European digital VLBI backends at the most stations had a valid bandwidth of only about 75 percent, the side channels were dropped out in loading the data into \textsc{aips}. With a reduction of the bandwidth, the task \textsc{accor} was performed to correct the cross-correlation amplitude.   

A priori amplitude calibration was performed with the system temperatures and the antenna gain curves. For cases where the telescope monitoring data was missing, nominal values of the system equivalent flux density in the EVN status table were used. The ionospheric dispersive delays were corrected according to a map of total electron content provided by Global Positioning System (GPS) satellite observations. Phase errors due to antenna parallactic angle variations were removed. After a manual phase calibration was carried out, the global fringe-fitting and the bandpass calibration were performed.

At 4.93 GHz, we first imaged the phase-referencing calibrator J1111$+$3252. We iteratively ran model fitting with point sources and self-calibration in \textsc{difmap} \citep[version 2.5e,][]{Shepherd1994}, fringe-fitting and self-calibration to remove its structure-dependent phase errors in \textsc{aips}. The calibrator has a single-side core-jet structure with a total flux density of 0.16$\pm$0.02~Jy. Its radio core, i.e. the jet base, has a peak brightness of 0.09$\pm$0.01~Jy~beam$^{-1}$ and its position was used as the reference point in the phase-referencing calibration. We also ran amplitude self-calibrations on the calibrator data and transferred the solutions to the target data. Owing to the limited $uv$ coverage of the short observations in particular on the long baselines, the deconvolution was performed by fitting the visibility data directly to some point source models in \textsc{difmap} to minimise the potential deconvolution errors of \textsc{clean}. 

\begin{table*}
\caption{The experiment setups of the used EVN observations. Columns give  (1) date, (2) frequency, (3) total time, (4)  data rate, (5) baseband filter, (6) project code, (7) observing mode and (8) participating stations (see Section~\ref{sec2-1} for the explanation on the telescope codes).}
\label{tab2}
\begin{tabular}{cccccccc}
\hline\hline
 Date       & $\nu_{\rm obs}$ & Time   & Rate   & Filter & Project & Observing mode              & Participating stations \\                              
            & (GHz)           & (h)    & (Mbps) & (MHz)  & Code    &                    &                        \\
\hline
2016 Mar 08 & 1.66            & 4.0    & 1024   & 16     & EY024B  & Disk recording VLBI  & Ro, Sr, Hh, Zc, Sv, Tr, Ur, T6, O8, Mc, Ef, Wb, Jb1 \\    
2016 Nov 16 & 4.93            & 2.5    & 2048   & 32     & RSY04   &  Real-time e-VLBI &  Ir, Hh, T6, Ys, O8, Nt, Mc, Ef, Wb, Jb2  \\
\hline
\end{tabular}
\end{table*} 

We performed about ten times iterations of the deconvolution and the self-calibration at 1.66 GHz. At each iteration, the new image model was used to solve for the residual systematic phase and amplitude errors in \textsc{aips}. When some significant ($\ga5\sigma$) positive features were found in the residual intensity map and the self-calibration failed to remove them, new model-fitting delta components were manually added to improve the fitting in \textsc{difmap}. Totally, there were 35 point-source components used at 1.66-GHz. The reliability of each main feature was also further verified by removing it and seeing the beam pattern around it with proper data weighting and tapering. As a consequence of no observations of the phase-referencing calibrator at 1.66~GHz, the image peak positions were used to align the images at 1.66 and 4.93~GHz. 

We also tried the standard \textsc{clean} algorithm with carefully applying data weighting and tapering in \textsc{Difmap}. Using the final self-calibrated data, we could make a consistent map while with a relatively high (about 1.3 times) noise level mainly in the on-source region. This is a known limitation of the \textsc{clean} algorithm because it generated a certain spurious structure in the form of spots or ridges as modulation on the broad features. Thus, we prefer the model-fitting technique to the \textsc{clean} algorithm.

\begin{table*}
\caption{Circular Gaussian model-fitting results of the major components found in \target{}. Columns give (1) name,  (2) observing frequency, (3) integrated flux density, (4--5) relative offsets in right ascension and declination with respect to the optical \textit{GAIA} position, (6) size, (7) brightness temperature, (8) radio luminosity. }
\label{tab3}
\begin{tabular}{cccccccc}
\hline\hline
Name   &   $\nu_{\rm obs}$ 
                             &         $S_{\rm obs}$     &  $\Delta\alpha\cos\delta$  &    $\Delta\delta$         & $\theta_{\rm size}$        &     $T_{\rm b}$     & $L_{\rm R}$     \\
             &  (GHz)  &        (mJy)                    &  (mas)                              &      (mas)                      & (mas)                            &   (K)                             & erg s$^{-1}$\\ 
\hline
N   &  1.66  & 0.26$\pm$0.03      &  $+1.13\pm$0.61       &  $+36.55\pm$0.56     &     11.6$\pm$1.3        &     1.0$\times10^6$  & 4.3$\times10^{38}$ \\
C   &  1.66  & 1.28$\pm$0.01      &  $-1.75\pm$0.02       &  $-1.73\pm$0.03      &      1.7$\pm$0.1      &    2.4$\times10^8$   & 2.1$\times10^{39}$ \\ 
S1  &  1.66  &  64.95$\pm$0.01    &  $-4.82\pm$0.01       &  $-16.78\pm$0.01     &      3.3$\pm$0.1    &    3.2$\times10^9$   & 1.1$\times10^{41}$  \\
S2  &  1.66  & 11.43$\pm$0.01     &  $-11.14\pm$0.01      &  $-16.00\pm$0.01     &      2.9$\pm$0.1     &    3.9$\times10^8$   & 1.9$\times10^{40}$ \\
W   &  1.66  & 0.29$\pm$0.01      &  $-19.83\pm$0.26      &  $-4.54\pm$0.29      &      4.6$\pm$0.6      &     7.2$\times10^6$  & 4.8$\times10^{38}$  \\
\hline
S1  & 4.93  & 9.54$\pm$0.07       &  $-5.13\pm$0.02       & $-16.95\pm$0.01      &      2.7$\pm$0.1      &  8.1$\times10^7$  &  4.6$\times10^{40}$ \\
S2  &  4.93 & 1.67$\pm$0.10        &  $-11.71\pm$0.17      & $-14.30\pm$0.11      &      5.3$\pm$0.1      &  3.5$\times10^6$  &  8.1$\times10^{39}$ \\
\hline
\end{tabular} \\ 
\end{table*}

\subsection{The VLA archive data}
\label{sec2-2} 

To study the compactness of \target{} on the kpc scales, we downloaded the archive visibility data of project AK0311 observed by the VLA at 1.4 and 8.4~GHz on 1992 December 20. The data were calibrated and provided by the image retrieval tool of the NRAO science data archive. The flux densities were reported in Table~\ref{tab1}.   

We also downloaded the raw visibility data of the VLA project AN0104. The project was observed by \citet{Nagar2003} at 15~GHz on 2002 February 23. According to the standard calibration steps recommended by the \textsc{aips} cookbook, we calibrated the data manually with a short solution interval of one minute. An image sensitivity of 0.16 mJy~beam$^{-1}$ was achieved. The quasar \target{} is detected with a signal to noise ratio (SNR) of 28, against the early report of non-detection \citep{Nagar2003}. 

\section{Results}
\label{sec3}

\subsection{Broad-band radio spectrum}
\label{sec3-1}
On the kpc scales, \target{} is unresolved in the VLA images. With a natural weighing, the VLA image sensitivities reach 0.13 mJy~beam$^{-1}$ at 1.43~GHz, 0.03 mJy~beam$^{-1}$ at 8.44 GHz, and 0.16 mJy~beam$^{-1}$ at 14.94~GHz. These VLA flux density measurements are also reported in Table~\ref{tab1}. 

There is no significant flux density variability observed so far. Our 1.43~GHz flux measurements are consistent with the results of the VLA surveys of the FIRST \citep[Faint Images of the Radio Sky at Twenty Centimeters,][]{Becker1995} and the NVSS \citep[NRAO VLA Sky Survey,][]{Condon1998}. The new 5.23~GHz flux measurement  \citep{Berton2018}  has no significant difference from that observed by the GBT about 30 years ago \citep{Gregory1996}. 

\begin{figure}
\centering
\includegraphics[width=0.46\textwidth]{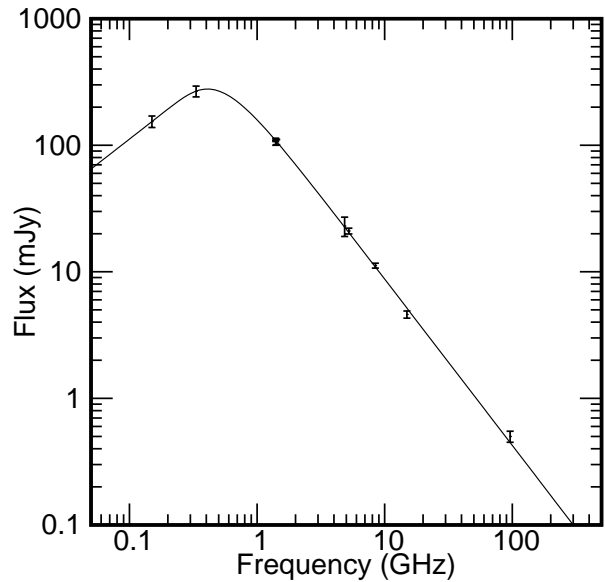}  \\
\caption{The broad-band radio spectra of \target{}. The data are listed in Table~\ref{tab1}. The grey curve shows the best-fit results of Equation~\ref{eq1}. }
\label{fig1}
\end{figure}

The broadband radio spectrum of \target{} in Fig.~\ref{fig1} shows a spectral shape similar to GHz-peaked spectrum sources \citep[e.g.][]{ODea1998} with a steep spectral slope at the high frequencies. To characterise the spectra, we also fit the data to a function similar to the synchrotron self-absorption model for the spherical homogeneous plasma. The function is
\begin{equation}
S_{\rm \nu}(\nu) = \frac{S_{\rm t}}{1-\exp(-1)} 
\left( \frac{\nu}{\nu_{\rm t}} \right)^{\alpha_{\rm r}} \left \{ 1 -\exp \left[ - \left( \frac{\nu}{\nu_{\rm t}} \right)^{\alpha - \alpha_{\rm r}} \right] \right \},  
\label{eq1} 
\end{equation}
where $S_{\rm t}$ is the flux density at the spectral turnover frequency $\nu_{\rm t}$, $\nu$ is the observing frequency, $\alpha$ is the spectral index in the optically thin region, and $\alpha_{\rm r}$ is the spectral index of the rising region, and $\alpha_{\rm r}=2.5$ in the synchrotron self-absoprtion mode for the spherical homogeneous plasma. The least-square fitting gives $S_{\rm t}=262\pm31$ mJy,  $\nu_{\rm t}=0.53\pm0.06$ GHz, $\alpha=-1.31\pm0.02$, $\alpha_{\rm r}=0.79\pm0.21$, and the reduced $\chi^2_{\rm r}=1.01$. 

\subsection{Parsec-scale radio morphology}
\label{sec3-2}
\begin{figure*}
\centering
\includegraphics[width=0.99\textwidth]{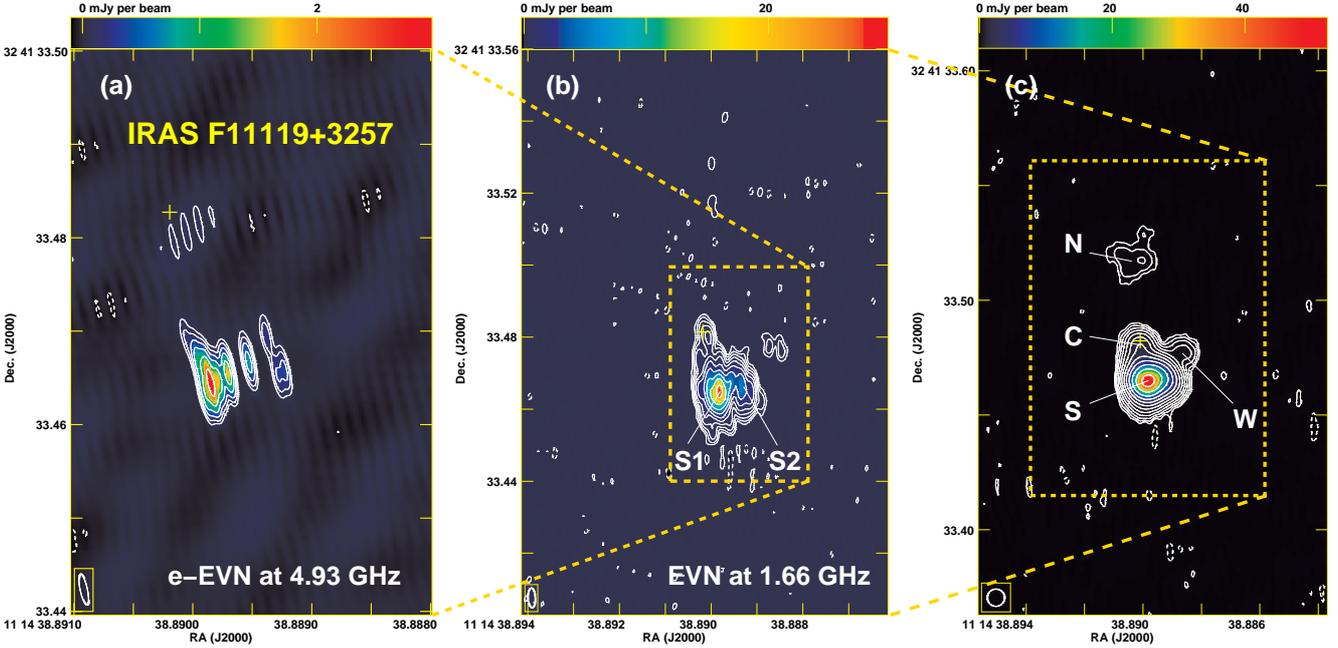}  \\
\caption{
The two-sided but significantly beamed jet observed with the EVN in the highly accreting quasar \target{}. The yellow cross marks the optical centroid measured by \textit{GAIA}. The first contours in all the images are at 3$\sigma$ level. (a) The high-resolution 4.93-GHz image is made with natural weighting. The full width at half maximum (FWHM) is 3.4~mas~$\times$~0.77~mas at 12.0 deg. The contours are 0.07~$\times$~($-$1, 1, 2, 4, ..., 32)~mJy~beam$^{-1}$. (b) The high-resolution 1.66-GHz image is made with uniform weighting. The FWHM is 5.4~mas~$\times$~2.0~mas at 2.7 deg. The contours are 0.045~$\times$~($-$1, 1, 2, ..., 512)~mJy~beam$^{-1}$.  (c) The high-sensitivity 1.66-GHz image is made with a circular Gaussian beam. The FWHM is 7.5~mas. The contours are 0.02~$\times$~($-$1, 1, 2, 4, ..., 2048)~mJy~beam$^{-1}$. }
\label{fig2}
\end{figure*}

All the EVN imaging results of \target{} are shown in Fig.~\ref{fig2}. The parsec-scale radio morphology shows a strong dependence on the observing frequency, the image resolution and sensitivity. Compared to the total flux densities predicted by our model, our VLBI images have restored nearly 100 percent at 1.66 GHz and about 50 percent at 4.93 GHz. 

The 4.93-GHz image in the left panel has the highest image resolution, up to 0.8 mas in the direction of East-west, because of the significant contribution of the Tianma 65-m radio telescope. Compared to the earlier imaging results from snap-shot observations with the Very Long Baseline Array (VLBA) at 2.3 GHz \citep{Petrov2013}, our image reveals the more details. Beside the peak component, there are a significant extension toward North and a few faint features with the decreasing brightness toward West. Because of the limited image sensitivity and $uv$ coverage, the extension toward West is coincidentally similar to the typical one-sided core-jet structures observed in flat-spectrum blazars \citep[e.g.][]{Cheng2018}. With respect to the phase-referencing calibrator, the image has quite precise position measurements. The optical centroid, reported by the \textit{Gaia} DR2 \citep{Brown2018}, is marked as a yellow cross (J2000, RA$=11^{\rm h}14^{\rm m}38\fs89019$, Dec.$=32\degr41\arcmin33\farcs4824$, $\sigma_{\rm ra}=\sigma_{\rm dec}=0.1$~mas, the astrometric excess noise: 0.4 mas). With respect to the optical centroid, the radio peak has an offset of about 18 mas. 

There are more faint features recovered at 1.66 GHz in the middle and right panels. As all the big antennas (Ef, Jb1, Ro, Sr, T6) are included, even a four-hour EVN observations still allows us to achieve an extremely high image sensitivity of 6.5~$\umu$Jy~beam$^{-1}$ (1$\sigma$) with naturally weighting. To clearly reveal these faint features, we convolved the source model with a larger circular beam of 7.5 mas. The low-resolution image in the right panel displays four relatively discrete components. According to their positions, they are marked as N, C, S and W. In the middle panel, component S is resolved into S1 and S2, and component N is not seen because of the relatively high image resolution. Both components S1 and S2 have faint extensions in almost all the directions. There also exists significant continuous radio emission connecting components S and C. To quantitatively describe these components, we also fit them with circular Gaussian models in \textsc{difmap} 2.5e. The least-square fitting results including the formal 1$\sigma$ uncertainties at the reduced $\chi^2_{\rm r}=1$ are summarised in Table~\ref{tab3}. The errors for the positions were truncated to 0.01 mas. The errors for the sizes were truncated to 0.1 mas. The empirical systematic uncertainties for $S_{\rm obs}$ and $L_{\rm R}$ are five percent.   

All the detected components in \target{} have a brightness temperature of $\geq1.0\times10^6$~K at 1.6 and 5 GHz. Component S1 has the highest brightness temperature reaching 3.2$\times10^9$~K at 1.6 GHz. The next to last column in Table~\ref{tab3} reports $T_{\rm b}$, estimated as \citep[e.g.][]{Condon1982},
\begin{equation}
T_{\rm b} = 1.22\times10^{9}\frac{S_\mathrm{obs}}{\nu_\mathrm{obs}^2\theta_\mathrm{size}^2}(1+z),
\label{eq2}
\end{equation}
where $S_\mathrm{obs}$ is the total flux density in mJy, $\nu_\mathrm{obs}$ is the observing frequency in GHz, $\theta_\mathrm{size}$ is the FWHM of the circular Gaussian model in mas, and $z$ is the redshift. 

\section{Discussion}
\label{sec4}
\subsection{No evidence for the star-formation activity}
All these VLBI-detected components originate in the AGN activity instead of the star-formation activity. Their radio emission are unlikely dominated by thermal emission of the star-forming activity because of their $T_{\rm B}\ga10^{6}$~K. It is also difficult to associate them with single young supernova or many supernovae remnants produced by the star-formation activity. The radio luminosity $L_{\rm R}=\nu L_\nu$ of each component is listed in the last column of Table~\ref{tab3}. A rarely-seen young supernova may reach a peak luminosity, $L_{\rm R}\sim$ $5\times10^{38}$~erg~s$^{-1}$ \citep[e.g.][]{Weiler2002}, comparable to $L_{\rm R}$ of the two faint components N and W, while fail to explain their extended structure. Additionally, because \target{} is not a starburst galaxy \citep{Komossa2006} like Arp 220 \citep[e.g.][]{Varenius2019}, the two components cannot be composed of overlapping emission from supernova remnants in the nuclear region.

\subsection{Non-detection of the flat-spectrum radio core}
\label{sec4-1}
None of these VLBI-detected components can be identified as the flat-spectrum radio core of \target{}  because of their extended structure ($\theta_{\rm size}\geq1.7$ mas) and optically thin spectra. Component C is relatively close to the optical centroid. However, it has a steep spectrum: with flux densities of 1.28 mJy at 1.66 GHz, $\le$0.07 mJy/beam (3$\sigma$) at 4.93 GHz, $\alpha\le-2.7$ ($S_\nu \propto \nu^{\alpha}$). The bright components S1 and S2 have a similar spectral index of $\alpha=-1.8$. The remaining components N and W are representative of the more extended structure and also have steep ($\alpha\le-1.3$) spectra. Moreover, according to the map of the velocity gradient of the CO(1--0) line emission of the host galaxy \citep{Veilleux2017}, there exists a hint for a small offset between the zero-velocity component and the continuum source, i.e. component S.   

The undetected radio core is most likely near component C. Firstly, this allows us to naturally explain the extended conical structure formed by components C and S at 1.66 GHz. Secondly, this is consistent with the \textit{Gaia} position. Generally, the \textit{Gaia} positions are strongly dependent on the assumption of the point source structure. As a luminous quasar, \target{} has a compact optical structure. This has been indirectly confirmed by the small astrometric excess noise ($1\sigma=0.4$ mas) in the Gaia position. Finally, according to a statistical analysis of systematic differences in the positions of \text{Gaia} DR2 with respect to VLBI \citep{Kovalev2017}, the separation $\sim$18 mas between the radio peak and the optical centroid is too large to be explained as a potential systematic position error ($\leq$10~mas).

The non-detection of the radio core allows us to set $3\sigma$ limits on its radio luminosity: $L_{\rm R} \leq3.4\times10^{38}$ erg~s$^{-1}$ at 4.93 GHz and $L_{\rm R}\le3.5\times 10^{37}$ erg~s$^{-1}$ at 1.66 GHz. 

The radio core may be intrinsically faint. This is also in agreement with the extremely steep spectra between $\nu_{t}=0.53\pm0.06$ and 96 GHz. At the low accretion rate state, there exists a correlation \citep[e.g.][]{Merloni2003} among the radio core luminosity at 5 GHz, the X-ray luminosity ($L_{\rm X}$) in the 2--10 keV and the black hole mass ($M_{\rm BH}$):
\begin{equation}
\log L_{\rm R} =(0.60^{+0.11}_{-0.11}) \log L_{\rm X} + (0.78^{+0.11}_{-0.09}) \log M_{\rm BH} + 7.33^{+4.05}_{-4.07} 
\label{eq3}
\end{equation}
According to the X-ray observations, $L_{\rm X}=3.8\times10^{42}$ erg~s$^{-1}$ \citep{Teng2010} at the high accretion rate state. Since the high-state $L_{\rm X}$ is most likely an upper limit, it would predict $L_{\rm R}\leq10^{38.6\pm0.88}$ erg~s$^{-1}$, comparable to our observational limits.

Another possible reason for the non-detection of the radio core may be its quenching after a transition from the low to high accretion rate states. By analogy with the unification evolution model for Galactic X-ray binaries \citep{Fender2004}, and assuming that the jet radiative properties are scale-free \cite[e.g.][]{Ruan2019}, the radio cores in AGN could be quenched when their accretion rates approach or exceed their Eddington accretion rates. Currently, the radio quiescence in the high state has been observed by \citet{Greene2006} in a sample of 19 low-mass galaxies with candidate massive black holes. 

\subsection{Constraints on the jet parameters}
\label{sec4-2}
These components in \target{} cannot be explained as the wide-angle outflows or winds. They have a ratio of $\frac{L_{\rm R}}{L_{\rm X}}$, at least one order of magnitude higher that the relation $\frac{L_{\rm R}}{L_{\rm X}}\sim10^{-5}$ observed by \cite{Laor2008} in the radio-quiet Palomar-Green quasar sample. Moreover, there is no diffuse bi-conical structure observed.    

We can identify bright components C, S and W as the approaching jet components with respect to the optical position. The non-detection of the radio core and the rapid drop of brightness and flux density in the inner component C indicates that they were from a relatively short-duration ejection event. Component N can be identified as a receding jet component or a relic jet because of its faintness. 

The large flux density of component S may be mainly attributable to the Doppler boosting while the starkly lower flux density of component N to Doppler de-boosting thus resulting in the large flux density ratio. This involves approaching and receding components as opposed to them being expanding shocks. This is as there are no high-brightness hot spots and edges, which can be brightened significantly by strong shocks, and, the linear polarisation (indicative of an interaction at the interface between the shock and surrounding medium) is very low ($3\sigma\leq1.3$ mJy) from the NVSS survey \citep{Condon1998}.

\begin{figure}
\centering
\includegraphics[width=0.48\textwidth]{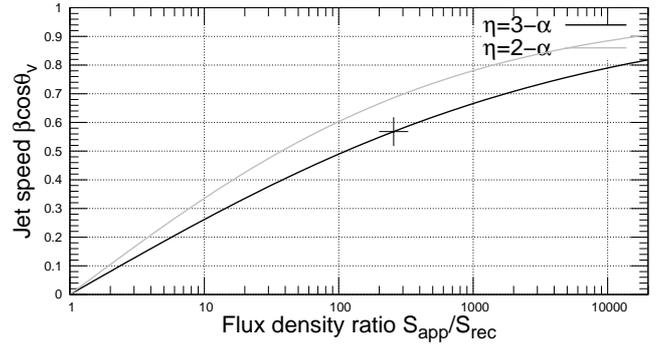}  \\
\caption{The relation between the flux density ratio $\frac{S_{\rm app}}{S_{\rm rec}}$ and the jet speed  $\beta\cos\theta_{\rm v}$ derived from Equation~\ref{eq4}. The cross marks the results of components S1 and N. }
\label{fig3}
\end{figure}

The apparent flux density ratio $R_{\rm flux}$ between approaching and receding jets emitting the same radio luminosity isotropically in their respective rest frames is \citep[e.g.,][]{Bottcher2012}
\begin{equation}
R_{\rm flux}=\frac{S_{\rm app}}{S_{\rm rec}} = \left(  \frac{1 + \beta\cos\theta_{\rm v}}{1-\beta\cos\theta_{\rm v}} \right) ^\eta,
\label{eq4}
\end{equation}
where $S_{\rm app}$ and $S_{\rm rec}$ are the flux densities of the approaching and receding components respectively,  $\beta$ is the intrinsic jet speed in unit of the light speed c, $\theta_{\rm v}$ is the jet viewing angle, $\eta=3-\alpha$ for a pair of discrete jet components and $\eta=2-\alpha$ for a continuous two-sided jet, and $\alpha=-1.31\pm0.02$ for our target. The relation is also plotted in Fig.~\ref{fig3}. The arm length ratio $R_{\rm arm}$ between approaching and receding components is also affected by the Doppler beaming effect, while much weaker than the flux density ratio ($R_{\rm arm}^{\eta}=R_{\rm flux}$). Because the approaching jet might bend significantly and the radio core is not identified, the arm length ratio is not available to constrain the jet parameters.

We can constrain the jet parameters $\beta$ and $\theta_{\rm v}$ using Equation~\ref{eq4}. If component S1 and N are a pair of approaching and receding ejecta and $\eta=3-\alpha$, we can derive a ratio of 246 and a constraint of $\beta\cos\theta_{\rm v} =0.57$c. The flux density ratio is also marked as a cross in Fig.~\ref{fig3}. The constraint gives $\beta\geq0.57$c and $\theta_{\rm v}\leq55$ degrees. Owing to the observed high ratio, the constraint on $\beta\cos\theta_{\rm v}$ is quite robust. It will not change significantly because the approaching jet may also include other much fainter components and $\eta=2-\alpha$. If component N is not the counter jet of component S1, the ratio will represent a lower limit and the limits on $\beta$ and $\theta_{\rm v}$ will remain valid. The small $\theta_{\rm v}$ would allow us to simply explain the apparent wide opening angle (about 60 degrees) and large change of the jet direction as a consequence of a projection effect.  

Our limit corresponds to an accretion disk inclination of $\theta_{\rm i}\leq55$ degrees assuming the jet is perpendicular to the accretion disk. As the lower limit of the jet speed is higher than that observed in the powerful X-ray outflows \citep{Tombesi2015} and only a maximum jet speed of upto 0.5$c$ is achievable for a radiative jet in a supercritical accretion disc \citep{Sadowski2015}, the jet components in \target{} may not be driven by radiation pressure alone. These jet components are not subject to highly relativistic beaming since they have an extended morphology, relatively low $T_{\rm B}$, and no high-energy $\gamma$-ray counterpart in the \textit{Fermi} LAT 8-Year Point Source Catalog \citep{Fermi2020}.

\target{} has the highest jet speed among the known highly-accreting objects. Super-Eddington accretions onto black holes are generally found in luminous AGN, TDEs, and ultra-luminous X-ray (ULX) sources. So far, all these super-critical accretion systems tend to have strong winds/outflows rather than relativistic jets. Among the TDEs, Swift J1644$+$57 had multi-band powerful non-thermal emission but an absence of a collimated synchrotron jet \citep{Yang2016a}. The only known TDE jet, found by \citet{Mattila2018} in Arp 299-B, had a relatively low jet speed of about 0.2$c$ on the pc-scale. Compared to the TDEs, the highly-accreting AGN have a much more stable radio emission. Currently, there are only a few targets imaged by the VLBI observations \citep[e.g.]{Giroletti2017, Yang2018}. This is also as most of them are radio quiet \citep[e.g., ][]{Panessa2007, Sikora2007} and have steep radio spectra \citep{Laor2019}, with a consequent lack of multi-epoch deep VLBI observations to constrain their jet speed. Additionally, extragalactic ULXs are rather faint for the VLBI observations to detect them \citep{Yang2016b}.      

\subsection{An unusual compact symmetric object}
\label{sec4-3}

\target{} may be identified as an uncommon compact symmetric object \citep[CSO, e.g.][]{Wilkinson1994, Readhead1996} from its two-sided jet morphology, compact size and exceptionally high flux density ratio. Most CSOs have faint radio cores and two-sided mini-lobes with projected sizes $\la$1~kpc and flux density ratios $\la10$ due to their jet speeds $\beta \la 0.9$ c and jet viewing angles $\theta_{\rm v} \ga 45\degr$ \citep[e.g.,][]{Owsianik1998, Polatidis2003, An2012b}. Compared to other CSOs \cite[e.g.,][]{Sokolovsky2011, An2012a}, its jet has a lower $T_{\rm b}$ owing to its steep spectrum and relatively extended structure. Moreover, it cannot be taken as a representative CSO because of its especially high flux density ratio.

According to the observed broad-band radio spectrum, \target{} is a Gigahertz-peaked spectrum (GPS) source \citep[e.g.][]{Stanghellini1998} or a compact steep-spectrum source \citep[CSS, e.g.][]{ODea1998}. Among the GPS source, CSOs are frequently found \citep[e.g.][]{Xiang2005}. According to the correlation relation between the linear size $l_{\rm jet}$ and the turnover frequency $\nu_{\rm t}$, $\log l_{\rm jet} = -1.54 \log \nu_{\rm t} - 0.32$,  observed in GPS and CSS sources \citep[][]{ODea1998}, the observed turnover frequency allows us to set a $3\sigma$ upper limit of 10 kpc for its linear size.  Using the lower limit of its jet speed (0.57 c), we can derive an upper limit on its kinematic age, $6\times10^5$ yr.  
If these jets were launched at the early low-accretion rate state, the age would set the first observational constraint on the time of the state transition of the accreting SMBH. Moreover, \target{} is also a low-power or low-luminosity ($L_{\rm R}\la10^{41}$~erg\,s$^{-1}$ at 1.4~GHz) compact radio source \citep{Giroletti2005}. It would require more energy injections for the compact source to become a large-scale Fanaroff-Riley type radio source \citep[e.g.][]{An2012b, Kunert2010}.

The turnover in the broad-band radio spectrum of \target{} might mainly result from synchrotron self-absorption. As the brightest component, component S1 has an average $T_{\rm b}$ from 8.1$\times$10$^7$ K at 5 GHz to 3.2$\times$10$^9$ K at 1.66 GHz.  Assuming that $\log T_{\rm b}$ and $\log \nu_{\rm obs}$ are linearly related, component S would have an average $T_{\rm b}$ of $1.3\times10^{11}$ K at $\nu_{\rm t}=0.53$ GHz. On the other hand, when a source is self-absorbed, the theoretical $T_{\rm b}$ in the emission rest frame is also $\sim10^{11}$ K at the peak frequency \citep{Readhead1994}. Thus, the inner part of component S1 might suffer significant synchrotron self-absorption. 

\section{Conclusions}
\label{sec5}
As an optically luminous quasar, \target{} hosts a supercritical accretion, the emission from which drives powerful X-ray outflows. Its radio spectrum between 0.15 and 96 GHz shows a peak at 0.53$\pm$0.06 GHz and a steep slope of $\nu^{-1.31\pm0.02}$ in the optically thin part. From the EVN observations at 1.66 and 4.93 GHz, the quasar displays a two-sided jet with a projected separation of about two hundred parsec. From the large flux density ratios between the approaching and receding jet components, we inferred that the jet has an intrinsic speed of $\geq$0.57$c$. This is higher than that observed in the X-ray winds and is thus unlikely to be driven by the radiation pressure alone. Among the known super-Eddington accretion systems, the jet in \target{} has the highest speed. Moreover, we identified \target{} as an unusual CSO with a jet viewing angle of $\leq$55 degrees, a kinematic age of $\leq6\times10^5$ yr, and a synchrotron self-absorbed radio spectrum. 
  
\section*{Acknowledgements}
\label{ack}
This work was partly supported by the SKA pre-research funding from the Ministry of Science and Technology of China (2018YFA0404603) and the Chinese Academy of Sciences (CAS, No. 114231KYSB20170003). 
% EVN
The European VLBI Network is a joint facility of independent European, African, Asian, and North American radio astronomy institutes. Scientific results from data presented in this publication are derived from the following EVN project codes: EY024B and RSY04. 
% VLA
The National Radio Astronomy Observatory is a facility of the National Science Foundation operated under cooperative agreement by Associated Universities, Inc.
% Gaia DR2
This work has made use of data from the European Space Agency (ESA) mission {\it Gaia} (\url{https://www.cosmos.esa.int/gaia}), processed by the {\it Gaia} Data Processing and Analysis Consortium (DPAC, \url{https://www.cosmos.esa.int/web/gaia/dpac/consortium}). Funding for the DPAC has been provided by national institutions, in particular the institutions participating in the {\it Gaia} Multilateral Agreement. 
% GMRT
We thank the staff of the GMRT that made these observations possible. The GMRT is run by the National Centre for Radio Astrophysics of the Tata Institute of Fundamental Research.
%%%%%%%%%%%%%%%%%%%%%%%%%%%%%%%%%%%%%%%%%%%%%%%%%%



%\bibliographystyle{mnras}

% My notes
% (1) <= 8 authors, list all of their names (Surname X.~Y.~Z.). 
% (2) > 8 authors, only list the first author and use ", et al.," for the rest authors
% (3) Remove "&"  between authors in case of two and multiple author paper in the references. 
% (4) Use "&" between the last two authors in case of citing two or three-author paper. No ","
% (5) Insert a space between initials with "~"

%%%%%%%%%%%%%%%%%%%%%%%%%%%%%%%%%%%%%%%%%%%%%%%%%%

%%%%%%%%%%%%%%%%% APPENDICES %%%%%%%%%%%%%%%%%%%%%

\appendix



%%%%%%%%%%%%%%%%%%%%%%%%%%%%%%%%%%%%%%%%%%%%%%%%%%

% Don't change these lines
\bsp	% typesetting comment
\label{lastpage}
\end{document}